\documentclass[a4paper]{jpconf}
\usepackage{graphicx}
\usepackage{amsmath,amsthm, amssymb, latexsym}
  \usepackage{tikz}
 \usetikzlibrary{arrows}
 \usetikzlibrary{decorations.markings}
 \usepackage{comment}
\begin{document}
\title{Measuring the Top Quark Mass using Kinematic Endpoints}

\author{Benjamin Nachman, on behalf of the CMS Collaboration}

\address{Laboratory for Elementary Particle Physics, Cornell University}

\ead{bnachman@cern.ch}

\begin{abstract}
We report on a simultaneous measurement of the top quark, W boson, and neutrino
masses in $t\bar{t}$ dilepton decays with $4.98 \mathrm{fb}^{-1}$ of data from the CMS experiment. The analysis is based on endpoint determinations in kinematic distributions. In addition to the unconstrained fit for three masses, the neutrino and W boson masses can be
constrained to standard values; in the maximally constrained fit, the top quark mass
is found to be $M_t = 173.9\pm0.9 \text{ (stat) }^{+1.2}_{-1.8} \text{ (syst)} $ GeV. 
\end{abstract}

\section{Introduction}

In many proposed extensions of the Standard Model, there is a new $\mathbb{Z}_2$ symmetry, like $R$ parity in Supersymmetry (SUSY), under which the new physics is odd and the known particles are even. New physics particles are thus produced in pairs and decays must terminate in a stable particle.  Such particles are candidates for dark matter and escape direct observation in the CMS detector~\cite{detector}.  We have developed a model independent method to simultaneously extract all unknown masses in a decay chain.  With the production of pairs of neutrinos, dileptonic $t\bar{t}$ resembles the generic new physics topology.   Thus, we test this method in data by simultaneously computing the top quark, W boson and neutrino masses.  Going beyond the general study of the technique, we also fix the $\mathrm{W}$ and neutrino masses to their known values and perform a measurement of the top quark mass.  This measurement is competitive in the dilepton channel and does not rely on Monte Carlo simulation.  More details are available in Ref.~\cite{ournote}.

\section{Kinematic Variables - $M_{T2}$}

\begin{figure}[h]
		\begin{center}
			\includegraphics[scale=0.35]{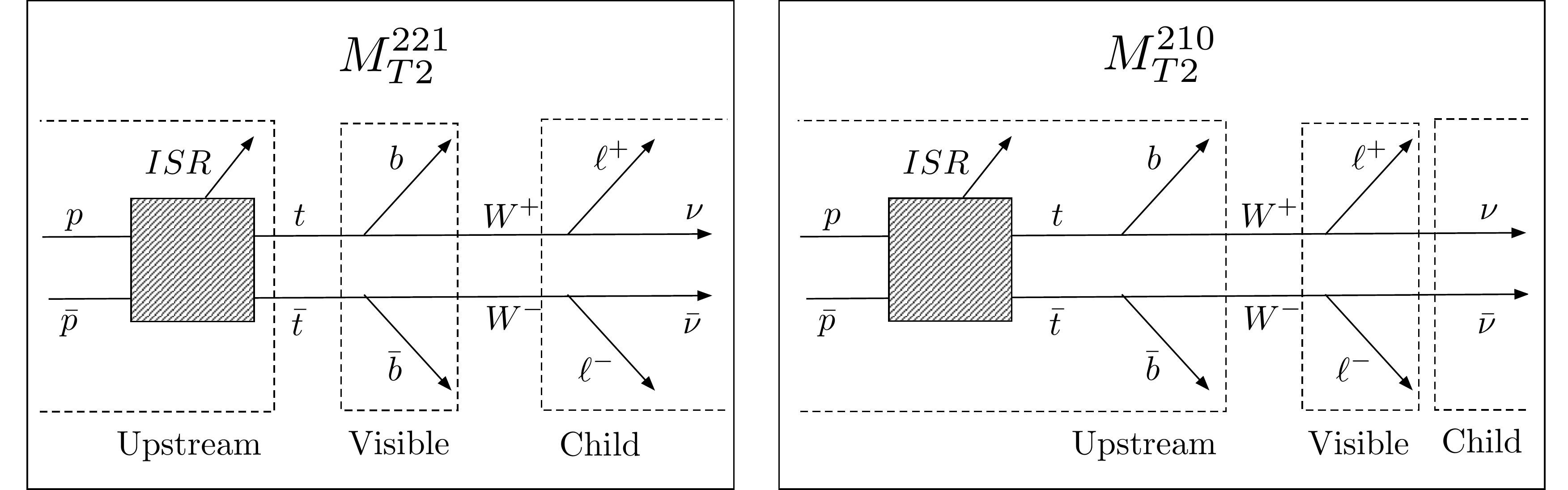}
          	\end{center}
	\vspace{-6mm}
		\caption{We partition the $t\bar{t}$ decay into subsystems.  For a third variable, we use the $bl$ invariant mass, which is in many cases the same as the third possible $M_{T2}$ variable we could form.}
		\label{subsyst}
\end{figure}

To combat the problem of underconstrained kinematics, we look at extreme kinematic features of distributions.  In particular, endpoints of distributions are determined and related to the underlying masses. This analysis uses the extension of transverse mass called $M_{T2}$~\cite{first}.   $M_{T2}$ is the minimum parent mass consistent with observed kinematics.  We have three unknowns, $M_\nu,M_W$, and $M_t$, so we partition the $t\bar{t}$ decay in order to create three mass variables~\cite{complete}.  Two $M_{T2}$ subsystem variables are combined with the $bl$ invariant mass to simultaneously extract the three masses.  Figure~\ref{subsyst} details the partitioning of the event topology for the $M_{T2}$ variables.  In general, the resulting variables depend on the magnitude of the transverse upstream momentum, $\vec{P}_T$~\cite{perp}.  To remove this dependence, we project all momenta perpendicular to $\vec{P}_T$ and compute $M_{T2\perp}$.  From~\cite{complete}, we have analytic formulae that relate distribution endpoints and underlying masses.

\section{2011 Data ($4.98/\mathrm{fb}$)}

\begin{figure}[h]
		\begin{center}
   	 \includegraphics[scale=0.26]{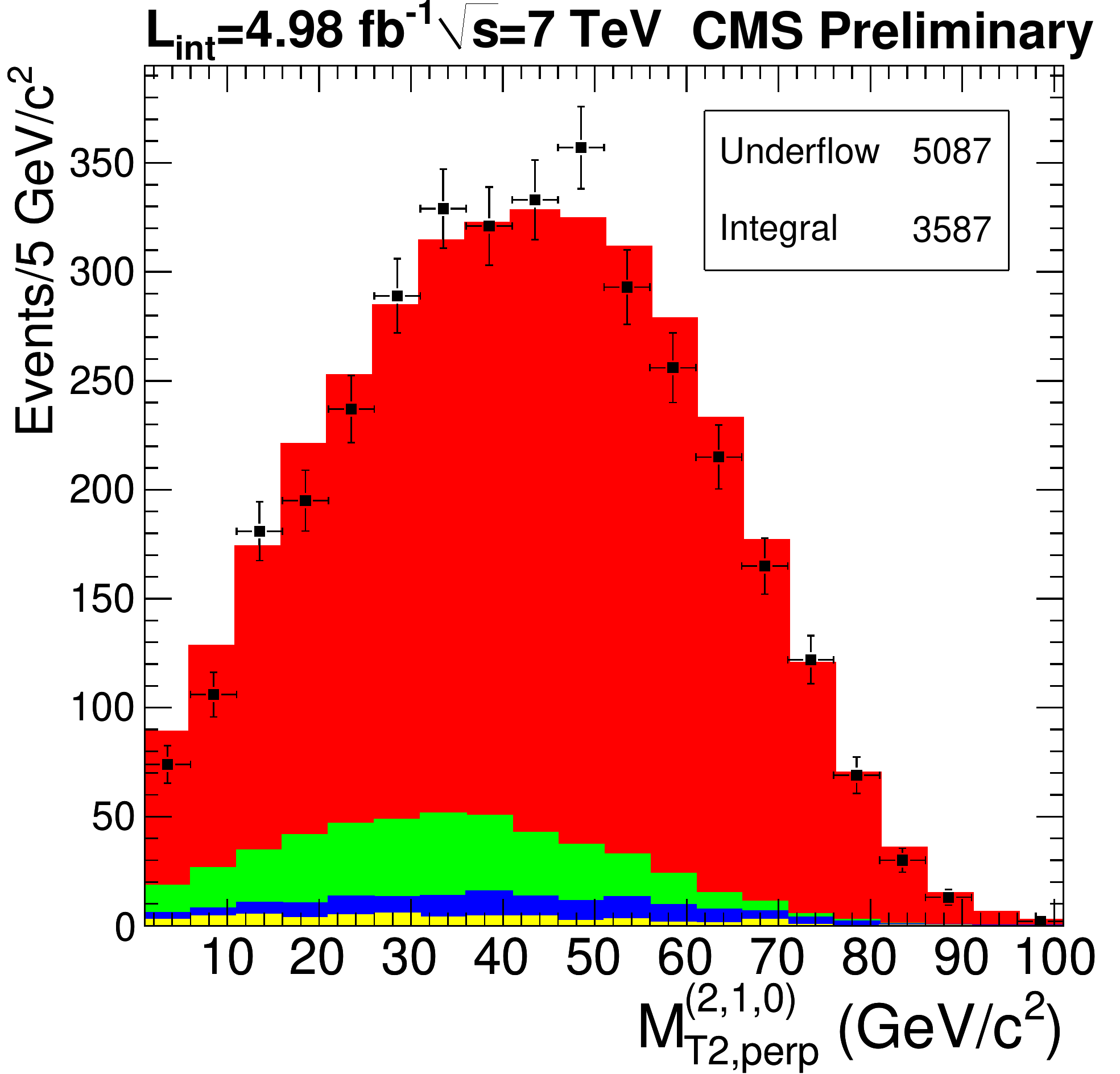}
    	 \includegraphics[scale=0.26]{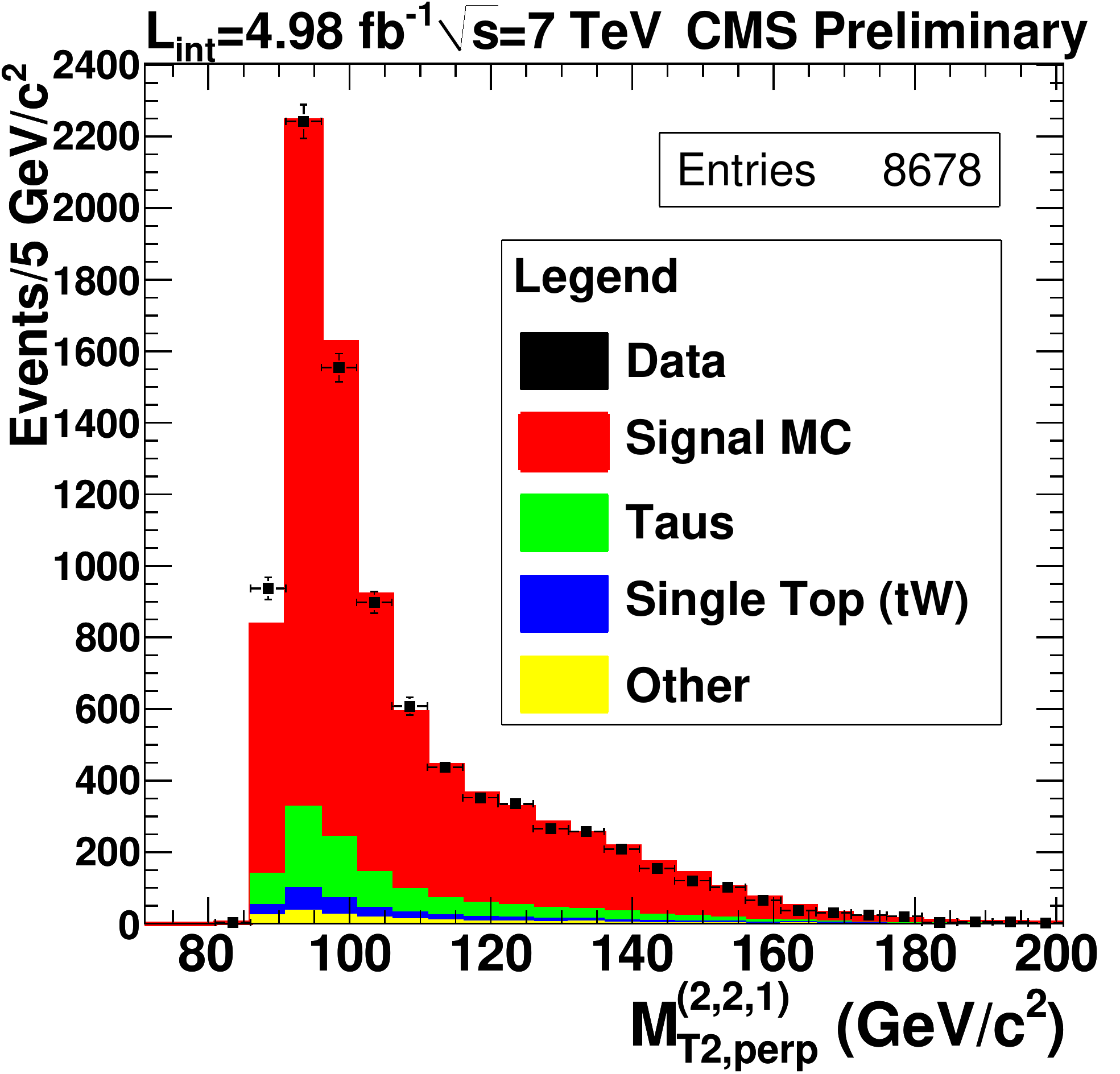}
        	\includegraphics[scale=0.26]{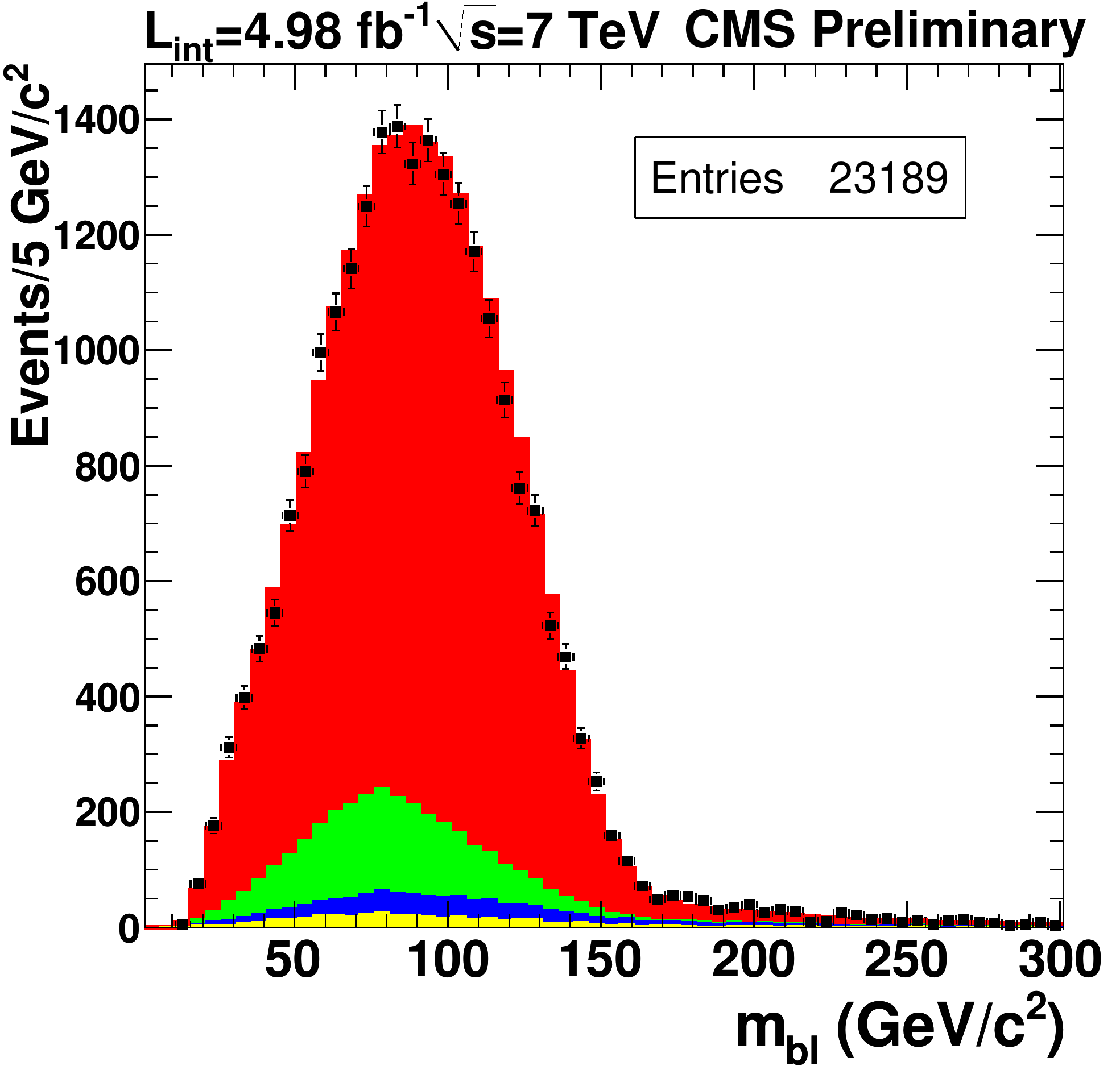}
          	\end{center}
	\vspace{-5mm}
		\caption{Monte Carlo overlaid on the 2011 data for the three measured kinematic distributions $M_{T2\perp}^{(210)}$, $M_{T2\perp}^{(221)}$, and $M_{bl}$.  The MC is luminosity normalized to the data with NLO cross sections.}
		\label{data}
\end{figure}

Monte Carlo (MC) is overlaid on the 2011 data to illustrate the approximate signal and background content of the distributions in Figure~\ref{data}. There is good agreement, but the signal MC is not used in the reported measurement.  From the known values of $M_t,M_W$, and $M_\nu$ the expected endpoints of $M_{T2\perp}^{(210)}$, $M_{T2\perp}^{(221)}$, and $m_{bl}$ are $80.4$, $173$, and $152.6$ $\mathrm{GeV}$ (respectively).

\section{Fitting Strategy}

We perform an unbinned maximum likelihood fit that takes into account event-by-event information such as resolution function shapes.  For $\text{\bf M}=(M_\nu^2,M_W,M_t)$\footnote{We are only sensitive to the square of the neutrino mass.} and data vector $u$ the likelihood is given by $\mathcal{L}(\text{\bf{M}})=\prod_a\prod_{i=1}^{N_{events}}\mathcal{L}_i^a(u_i|\text{\bf{M}})$ for $a\in\{M_{T2}^{210},M_{T2}^{221},M_{bl}\}$ and 

		\begin{align}
\text{ $\mathcal{L}_i^a(x_i|x_{\max})=$}&\text{$\alpha\int S^a(y|x_{\max}^a)\mathcal{R}_i^a(x_i-y)dy+(1-\alpha)B^a(x_i)$},
		\end{align}
		
\noindent where $\alpha$ is the signal normalization, $\mathcal{R}$ is a function that incorporates event-by-event resolutions, $B$ is the background estimation and $S$ is the signal shape.  This last quantity is approximated as linear decent into the endpoint in a region near the a priori kinematic maximum.

\section{Background Estimation}

\begin{figure}[h!]
		\begin{center}
   	 \includegraphics[scale=0.3]{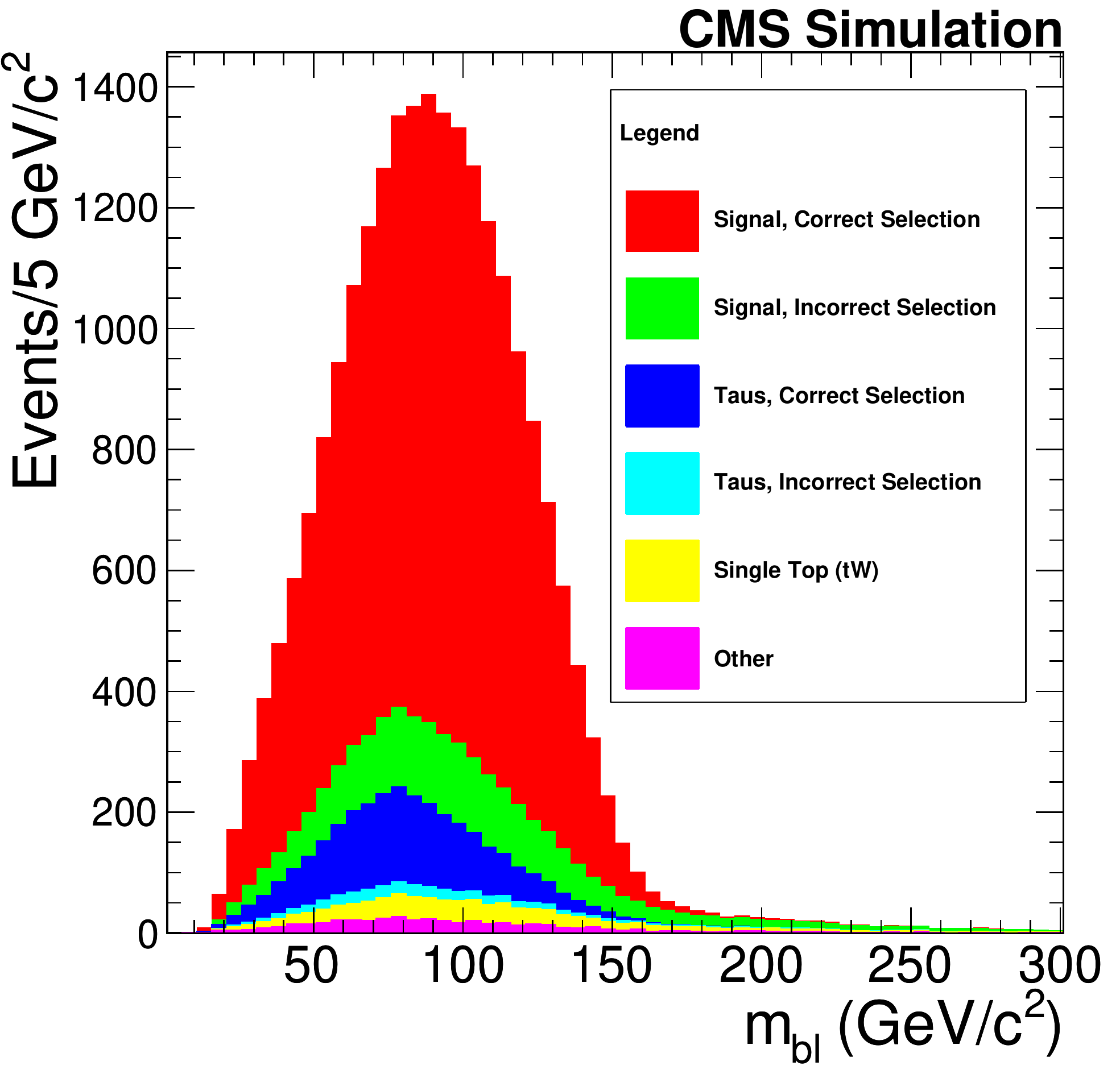}
    	 \includegraphics[scale=0.3]{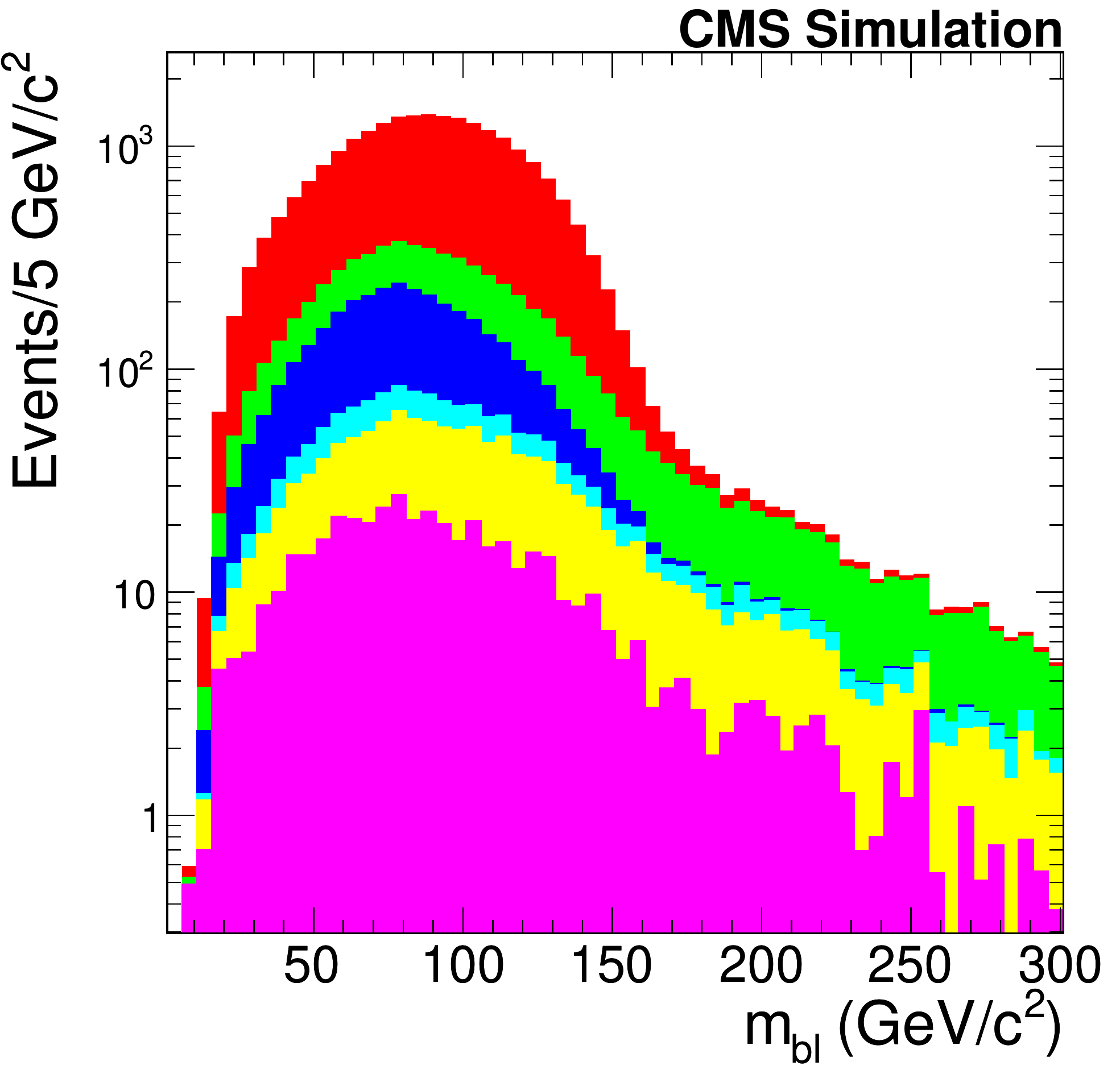}
	 \vspace{-4mm}
		\caption{Monte Carlo composition of the $M_{bl}$. Note that the dominant source of events beyond the endpoint are due to signal events in which the wrong jets were chosen as the $b$ jets.}
		\label{beyond}
		\end{center}
\end{figure}

The dominant source of background is combinatorial.  In particular, the largest contribution to this background is from $b$ jet selection, which is not $100\%$ pure.  Non $t\bar{t}$ events do not obey the same bounds as $t\bar{t}$.  Figure~\ref{beyond} shows the composition of events in the invariant mass distribution and makes clear that the mis-tag background dominates the events beyond the endpoint (153 GeV).  We model this dominant background by using a control sample in which we have inverted our $b$ tagging.  We then use a non-parametric estimation (adaptive kernel density estimation) of the shape which enters as $B^a$ in Eq. 6.  


\section{Uncertainty Estimation}

 The usual methods for statistical uncertainty estimation are not valid because they do not account for fluctuations in the background shape.  We thus use bootstrapping to estimate our statistical uncertainty.  By recomputing the fit to the control region for each bootstrap sample, we can take into account background fluctuations by taking the central value and width of the distribution of masses evaluated from the ensemble of bootstrapped samples.

We estimate our systematic uncertainties by inserting variations into our fit and re-evaluating the result.  Table \ref{t:results} summarizes the contributions to the total uncertainty.  Except for color reconnection, all the variations are applied to the data.

\begin{table}[htdp] 
\begin{center}
\begin{tabular}{|c|c||c|c|}
\hline
{Systematic}     &  {Uncertainty ($\mathrm{GeV}$)} &{Systematic}     &  {Uncertainty ($\mathrm{GeV}$)} \\[0.5ex]\hline
{Jet Energy Scale} & {$+0.5$}\hspace{5mm}{$-1.4$}   &{Fit Range } & {$\pm 0.6$}      \\[0.5ex]\hline
{ Jet Energy Resolution} &  {$\pm 0.5$}   &{ Background Modeling} &  {$\pm 0.5$}   \\[0.5ex]\hline
{ Pileup} & { neligible}     &{ Color Reconnection} & {$\pm 0.6$}     \\[0.5ex]\hline
\end{tabular}

\end{center}
\vspace{-4mm}
\caption{Systematic Uncertainties.  Except for color reconnection, the variations used in determining the uncertainties are applied to data.  The total uncertainty is $+1.2$, $-1.8$ GeV.}
\label{t:results}
\end{table}

\section{Validation}

\begin{figure}[h!]
		\begin{center}
	 \includegraphics[scale=0.3]{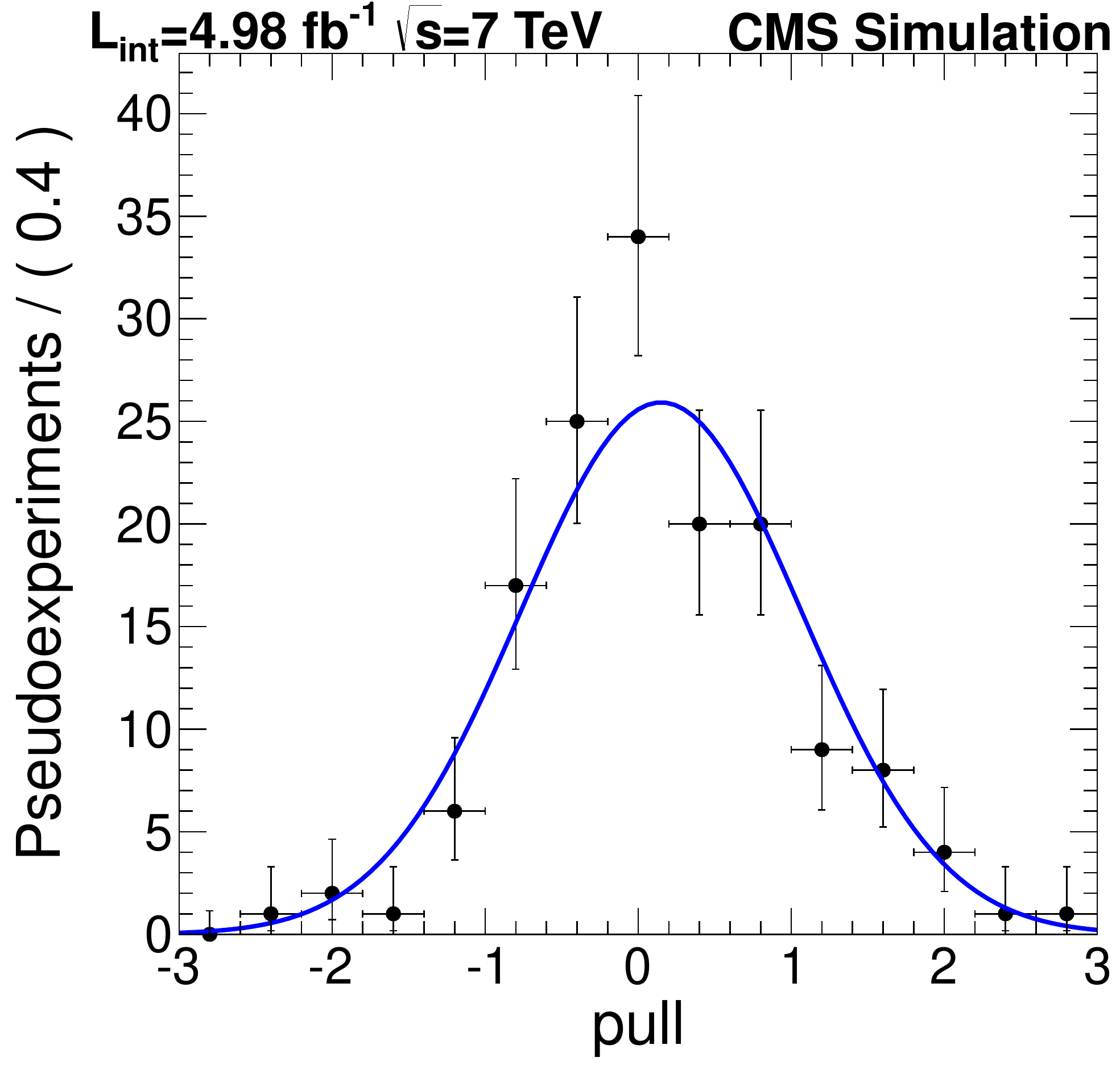}
	  \includegraphics[scale=0.3]{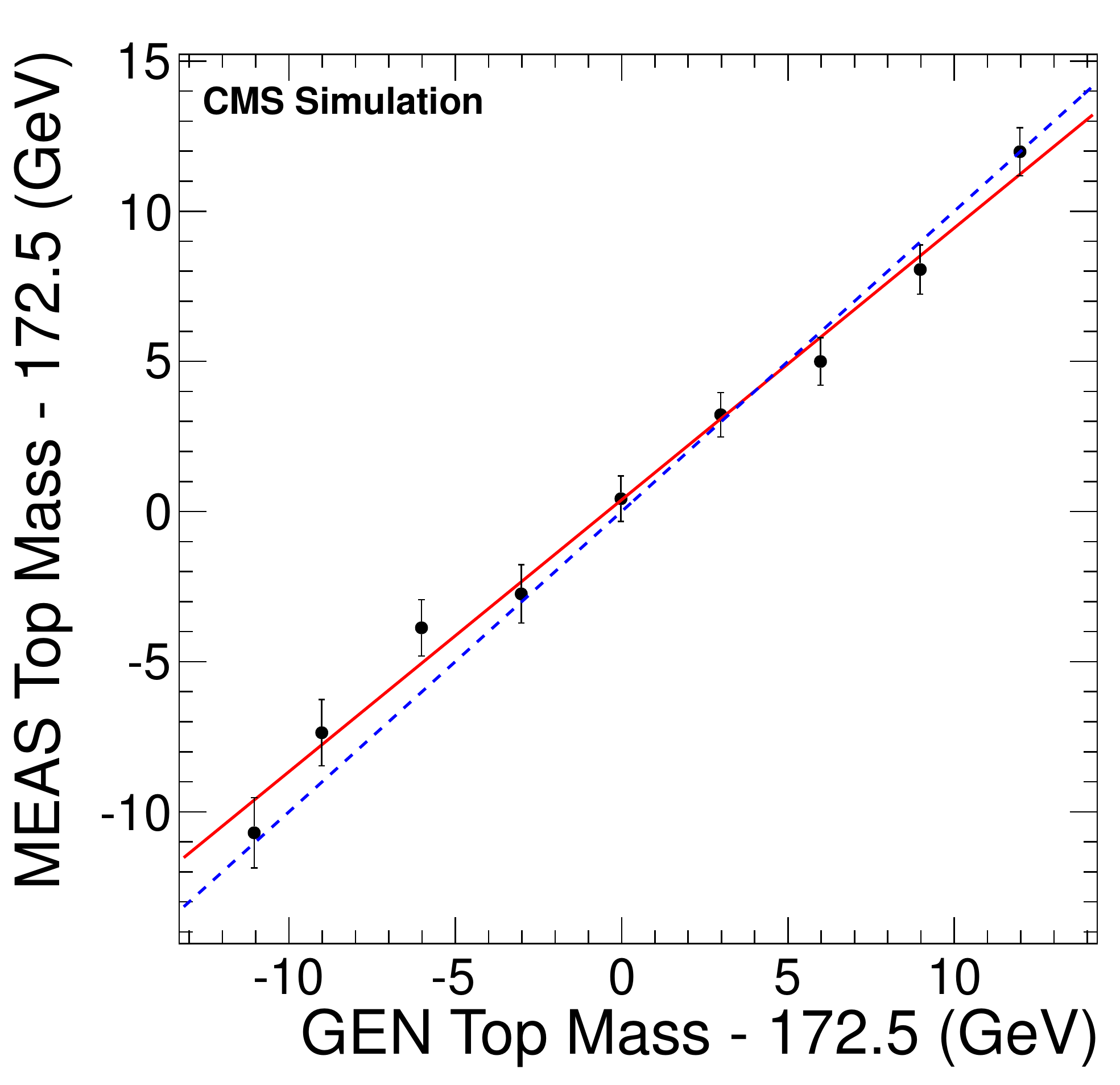}
	  \vspace{-4mm}
	  \caption{The left plot shows the pull distribution, with a mean of $0.15\pm 0.19$ and a standard deviation of $0.92\pm 0.06$.  The right plot shows the spread in measured top masses as a function of input top mass.  The best fit line is shown in  red (slope $0.91\pm 0.4$, intercept $0.34\pm 0.3$) and the line of unit slope is in blue.  The blue line agrees with the fit at $\chi^2/\mathrm{ndf}=10.7/9$.}
	  \label{valid}
	\end{center}
	\end{figure}

As a cross check of our analysis, we perform our fit on many pseudo-experiments (PE) of simulated data.  For each PE, define $\mathrm{Pull}=(m_{\mathrm{true}}-m_{\mathrm{meas}})/\sigma_{\mathrm{meas}}$.  If our procedure is unbiased and correctly produces the uncertainty, then the distribution of the pull should be a standard normal.  We can also probe potential bias by repeating this exercise for various values of $m_{\mathrm{true}}$.  As shown in Figure~\ref{valid}, both of these tests are consistent with an unbiased measurement with correct statistical uncertainty.

\section{Results}

We present three sets of results, corresponding to three levels of constraints on the masses of the neutrino and $\mathrm{W}$ boson.  The set of masses are shown in Table \ref{t:results2}.  In the case where both the neutrino and W boson masses are constrained, we compute our most precise top mass measurement: $M_t = 173.9\pm0.9 \text{ (stat) }^{+1.2}_{-1.8} \text{ (syst)} $ GeV.  Due to its minimal dependence on MC, this result has many complementary (systematic) uncertainties to other measurements and so can help reduce the world uncertainty.  However, in practice this is non-trivial as there is ambiguity on the order of $\Lambda_{QCD}$ as to what this analysis measures and what traditional MC calibrated approaches compute.  As the precision of top mass measurements improves, the top quark community will need to think carefully about the interpretation of the quantities being measured.

\begin{figure}[h!]
 \begin{center}
    \includegraphics[scale=0.65]{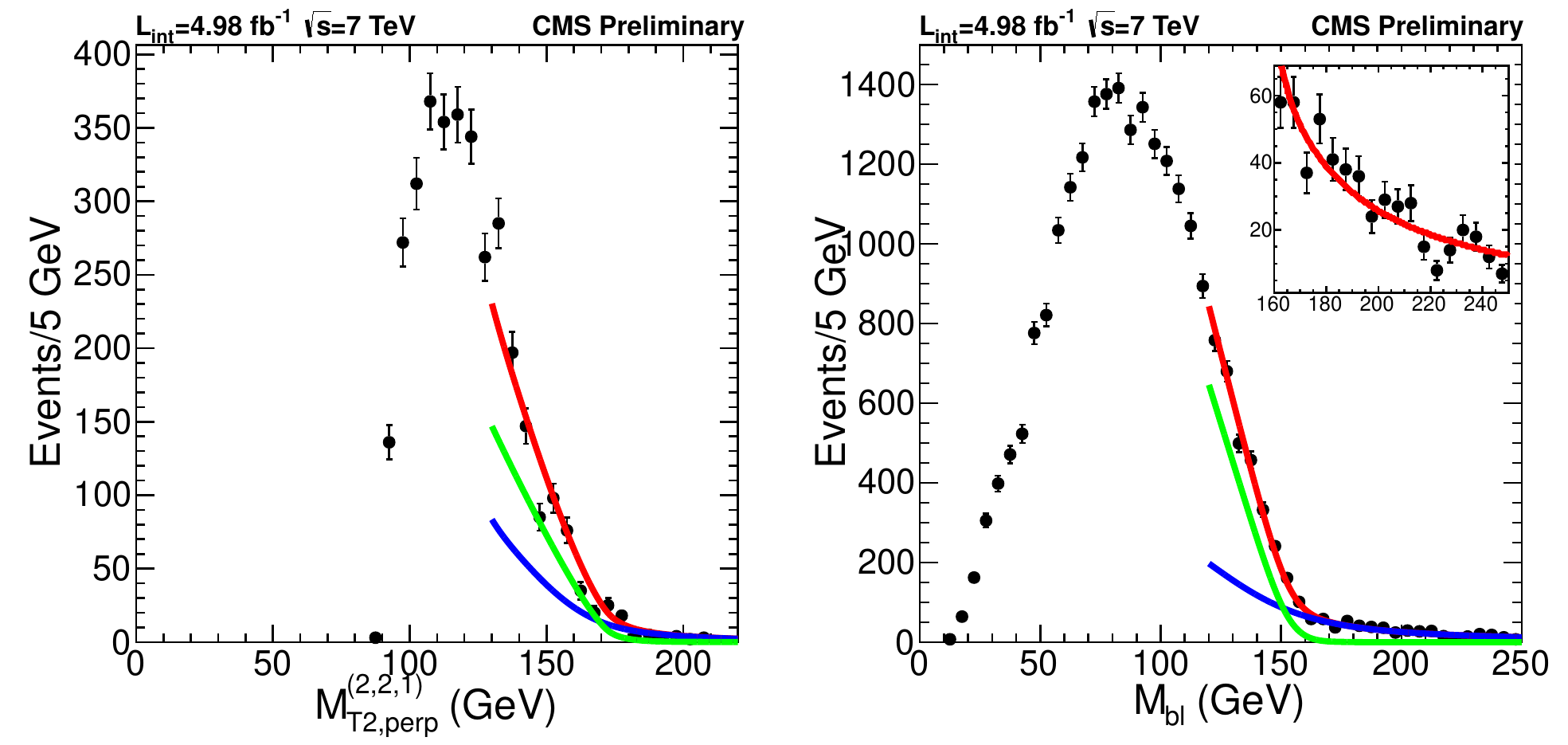}
    \end{center}
    \vspace{-7mm}
	  \caption{The fit plot showing the data and fit function for the case in which the neutrino and $W$ boson masses are fixed.  The displayed fit and data are from a particular bootstrapped sample.  The blue line is the background estimation from the control sample, the green line is the signal fit function and the red is their sum.  The top right of the $m_{bl}$ plot shows the endpoint region to illustrate the fit to the data.}
	\end{figure}
	
\begin{table}[htdp] 
\begin{center}
\begin{tabular}{|c|c|c|c|}
\hline&\multicolumn{3}{|c|}{{ Constraint}}\\
\cline{2-4}
\hline&&\multicolumn{2}{|c|}{{ $m_\nu=0$}}\cr
\cline{3-4}\rule{0pt}{3ex} 
~{ Fit Quantity}~    & ~~ { None} ~~& ~~~ & ~~{ $m_W=80.4$}~~ \\[1ex]\hline
\rule{0pt}{3ex} { $m_\nu^2$  (GeV$^{2}$) }& { $-556\pm473\pm600$} & { 0}  & { (0)}             \\[1ex]\hline
\rule{0pt}{3ex} { $m_W$ (GeV)}         & { $72\pm 7\pm 9$} & { $80.7\pm 1.1\pm 1$} & { (80.4)}       \rule{0pt}{3ex} \\[1ex]\hline
\rule{0pt}{3ex} { $m_t$ (GeV)}         & { $163\pm 10\pm 11$} & { $174.0\pm 0.9\pm 2$} & $173.9\pm 0.9^{+1.2}_{-1.8} $ \\[1ex]\hline
\end{tabular}
\end{center}
\label{t:results2}
\vspace{-4mm}
\caption{Results for the various levels of constraints.  Uncertainties are first statistical and then systematic.}
\end{table}

\section{Acknowledgements}

This work was partly under NSF support, PHY-0970024.


\section*{References}
\begin{thebibliography}{9}
\bibitem{detector}
CMS Collaboration, The CMS experiment at the CERN LHC, JINST 3 (2008) S08004.
\bibitem{ournote}CMS Collaboration, Mass determination in the $t\bar{t}$ system with kinematic
endpoints, CMS Physics Analysis Summary CMS-PAS-TOP-11-027 (2012).
\bibitem{first} C. G. Lester and D. J. Summers, ÒMeasuring masses of semi-invisibly decaying particle pairs produced at hadron collidersÓ, {\it Physics Letters B} 463 (1999), no. 1, 99-103.
\bibitem{complete} M. Burns, K. Kong, K. T. Matchev et al., ÒUsing Subsystem MT2 for Complete Mass Determinations in Decay Chains with Missing Energy at Hadron CollidersÓ, {\it JHEP} 0903 (2009) 143.
\bibitem{perp} P. Konar, K. Kong, K. T. Matchev et al., ÒSuperpartner Mass Measurement Technique using 1D Orthogonal Decompositions of the Cambridge Transverse Mass Variable MT2Ó, {\it Phys.Rev.Lett.} 105 (2010) 051802.
\end{thebibliography}
\end{document}